\documentclass[twoside]{article}
\usepackage{fleqn,espcrc2}

\def\JHEP #1 #2 #3{{\sl J. High. Energy. Phys.} {\bf#1} (#2) #3}

\def\PRL #1 #2 #3{{\sl Phys. Rev. Lett.} {\bf#1} (#2) #3}

\def\NPB #1 #2 #3{{\sl Nucl. Phys.} {\bf B#1} (#2) #3}

\def\NPBFS #1 #2 #3 #4{{\sl Nucl. Phys.} {\bf B#2} [FS#1] (#3) #4}

\def\CMP #1 #2 #3{{\sl Commun. Math. Phys.} {\bf #1} (#2) #3}

\def\PRD #1 #2 #3{{\sl Phys. Rev.} {\bf D#1} (#2) #3}

\def\PLA #1 #2 #3{{\sl Phys. Lett.} {\bf #1A} (#2) #3}

\def\PLB #1 #2 #3{{\sl Phys. Lett.} {\bf #1B} (#2) #3}

\def\JMP #1 #2 #3{{\sl J. Math. Phys.} {\bf #1} (#2) #3}

\def\PTP #1 #2 #3{{\sl Prog. Theor. Phys.} {\bf #1} (#2) #3}

\def\SPTP #1 #2 #3{{\sl Suppl. Prog. Theor. Phys.} {\bf #1} (#2) #3}

\def\AoP #1 #2 #3{{\sl Ann. of Phys.} {\bf #1} (#2) #3}

\def\PNAS #1 #2 #3{{\sl Proc. Natl. Acad. Sci. USA} {\bf #1} (#2) #3}

\def\RMP #1 #2 #3{{\sl Rev. Mod. Phys.} {\bf #1} (#2) #3}

\def\PR #1 #2 #3{{\sl Phys. Reports} {\bf #1} (#2) #3}

\def\AoM #1 #2 #3{{\sl Ann. of Math.} {\bf #1} (#2) #3}

\def\UMN #1 #2 #3{{\sl Usp. Mat. Nauk} {\bf #1} (#2) #3}

\def\FAP #1 #2 #3{{\sl Funkt. Anal. Prilozheniya} {\bf #1} (#2) #3}

\def\FAaIA #1 #2 #3{{\sl Functional Analysis and Its Application} {\bf

#1} (#2) #3}

\def\BAMS #1 #2 #3{{\sl Bull. Am. Math. Soc.} {\bf #1} (#2)

#3} \def\TAMS #1 #2 #3{{\sl Trans. Am. Math. Soc.} {\bf #1} (#2) #3}

\def\InvM #1 #2 #3{{\sl Invent. Math.} {\bf #1} (#2) #3}

\def\LMP #1 #2 #3{{\sl Letters in Math. Phys.} {\bf #1} (#2) #3}

\def\IJMPA #1 #2 #3{{\sl Int. J. Mod. Phys.} {\bf A#1} (#2) #3}

\def\AdM #1 #2 #3{{\sl Advances in Math.} {\bf #1} (#2) #3}

\def\RMaP #1 #2 #3{{\sl Reports on Math. Phys.} {\bf #1} (#2) #3}

\def\IJM #1 #2 #3{{\sl Ill. J. Math.} {\bf #1} (#2) #3}

\def\CMP #1 #2 #3{{\sl Commun. Math. Phys.} {\bf #1} (#2) #3}

\def\APP #1 #2 #3{{\sl Acta Phys. Polon.} {\bf #1} (#2) #3}

\def\TMP #1 #2 #3{{\sl Theor. Mat. Phys.} {\bf #1} (#2) #3}

\def\JPA #1 #2 #3{{\sl J. Physics} {\bf A#1} (#2) #3}

\def\JSM #1 #2 #3{{\sl J. Soviet Math.} {\bf #1} (#2) #3}

\def\MPLA #1 #2 #3{{\sl Mod. Phys. Lett.} {\bf A#1} (#2) #3}

\def\JETP #1 #2 #3{{\sl Sov. Phys. JETP} {\bf #1} (#2) #3}

\def\JETPL #1 #2 #3{{\sl  Sov. Phys. JETP Lett.} {\bf #1} (#2) #3}

\def\PHSA #1 #2 #3{{\sl Physica} {\bf A#1} (#2) #3}

\def\CQG #1 #2 #3{{\sl Class. Quantum Grav.} {\bf #1} (#2) #3}

\def\SJNP #1 #2 #3{{\sl Sov. J. Nucl. Phys. (Yadern. Fiz.)} {\bf #1} (#2) #3}
\def\FdP #1 #2 #3{{\sl Fortschr. Phys.} {\bf A#1} (#2) #3}
\def\SSC #1 #2 #3{{\sl Solid  State Commun.} {\bf #1} (#2) #3}

\def\a{\alpha}
\def\b{\beta}
\def\g{\gamma}
\def\d{\delta}
\def\e{\varepsilon}
\def\r{\varrho}
\def\k{\kappa}
\def\l{\lambda}
\def\L{\Lambda}
\def\s{\sigma}
\def\S{\Sigma}
\def\Th{\Theta}

\def\Om{\Omega}

\def\t{\tau}
\def\vp{\varphi}
\def\vs{\varsigma}
\def\g{\gamma}

\newcommand{\f}[1] {(\ref{#1})}

\newcommand{\AmS}{{\protect\the\textfont2
  A\kern-.1667em\lower.5ex\hbox{M}\kern-.125emS}}
\title{ Strings in a space with tensor central charge coordinates}
\author{ A.A. Zheltukhin\thanks{On leave of absence from Institute for Theoretical Physics, NSC  Kharkov Institute of Physics 
and Technology, 61108, Kharkov, Ukraine.}\thanks{E-mail: aaz@physto.se , zheltukhin@kipt.kharkov.ua } \thanks{Talk given at D.V. Volkov Memorial Conference,
 July  25-29, 2000,  Kharkov.}
and
U. Lindstr\"om \address{Institute of Theoretical Physics, University of Stockholm, Box 6730, S-11385 Stockholm, Sweden}\thanks{E-mail: ul@physto.se }}

\begin{document}

\begin{abstract}
New string models in $D=4$ space-time extended   by tensor central charge coordinates $z_{mn}$ are constructed.
We use the $z_{mn}$ coordinates to generate  string tension using a minimally extended string action linear in $dz^{mn}$.
It is shown that the presence of $z_{mn}$ lifts  the light-like character of the tensionless string worldsheet and the degeneracy of its induced metric. We analyse the  equations of motion  and find a solution of the string equations in the generalized D=(4+6)-dimensional space $X^{\cal M}=(x^{m},z^{mn})$ with  $z_{mn}$ describing a spin wave process. 
A supersymmetric version of the proposed model is formulated.

\vspace{1pc}
\end{abstract}

\maketitle

\vspace{10mm}

\section {Introduction}

   Development of noncommutative geometry ideas \cite{alb},
\cite{codos}, \cite{dohul}, \cite{seiwi}  have resulted in the discovery of Noncommutative Open String theory \cite{seisut}, \cite{gomms},
 \cite{gomss}. An interesting feature of the NCOS theory is the  appearence of a critical value of the  electric field \cite{gukp} for which the effective string tension becomes equal to zero.
 So, the  question arises as  to  the connection of this theory to  the standard theory of tensionless strings  \cite{sch},\cite{kl} and branes \cite{zb}. This question is further motivated by the results of papers \cite{lir},  \cite{lizz}, where a Born-Infeld Dp-brane action was constructed and studied in the limit of zero tension. It was shown in \cite{lizz} that the generalized gauge invariant Born-Infeld 2-form ${\cal F} = B+ ({\g\over2\pi\a^{\prime}})^{-1}F$ splits into two mutually orthogonal parts and one of them lies in the tangent plane spanned by the vielbein components $e^{0}_{i}$ and $e^{1}_{i}$ of tensionless Dp-brane. For the case $B=0$ the electric field  $E$ of the Dp-brane becomes constant
\footnote{Here ${\g}$ is a parameter characterizing the background metric \cite{zr}.}
 $E= {\g\over\pi\a^{\prime}}$ and directed along $e^{1}_{i}$.
 This observation is related to the character of the dynamics of tensionless Dp-brane  \cite{lir} which is reduced to the dynamics of an  effective string stretched along the $e^{1}_{i}$ direction.
These two facts lead to the conclusion that the tension of an effective open string (and Dp-brane) embedded into the Dp-brane is entirely compensated by the critical electric field  $E= {\g\over\pi\a^{\prime}}$ pulling apart the charges attached to the string's ends.
Therefore, the question about  the  noncommutative nature of the space-time coordinates in string/M- theory has to be intrincically related to  the question on the nature of the tension of extended objects.
 
Some  mechanisms for the generation of tension  have been discussed in  \cite{zi}, \cite{town1}, \cite{belt}, \cite{haliu}, \cite{gz} and they imply that tension is created by the  interactions of a  tensionless string/brane with some additional fields or coordinates. Selfconsistency and completeness of this picture suggest  that the additional field or coordinate should be an  object intrinsic to  string/M-theory.

From this point of view the coordinates corresponding to the tensorial central charges $Z_{m_{1},m_{2},...,m_{p}}$ are interesting. The point is that the central charges are connected with p-branes as solitonic solutions of supergravity equations  and  modify  the Poincare superalgebra to the form \cite{hupol}, \cite{azgit}, \cite{dust}, \cite{dafre}, \cite{holpro}, \cite{ziz}
\begin{eqnarray*}
\{ Q_{\a}, Q_{\b} \}=(\g^{m}C^{-1})_{\a\b}P_{m}\\
+\sum_{p}\left(\g^{m_{1},m_{2},...,m_{p}}C^{-1}\right)_{\a\b}Z_{m_{1},m_{2},...,m_{p}} 
\end{eqnarray*}
A non-trivial role of the new coordinates corresponding to the p-form generators  $ Z_{m_{1},m_{2},...,m_{p}}$ was emphasized in 
 \cite{curt}, \cite{gren}, \cite{hew}, \cite{eis} and advanced further in 
\cite{sieg}, \cite{bes}, \cite{se}, where  new superalgebras were constructed.
The string/M-theory approach  \cite{wit} considers the extended  superalgebras \cite{town2} of Dp-branes \cite{pol1} to be  connected with the non-perturbative string dynamics.

 Moreover, the  central charge carried by  the BPS brane/string  preserving $1/2$ of the N=1 supersymmety \cite{azgit} appears in QCD and  is associated with the domain wall created by the gluino condensate  \cite{ds}  (see also \cite{gs} and refs. there) and  spontaneous breakdown of the discrete chiral symmetry \cite{vy}. Recently it was shown that the intersection of domain walls leads to creation of configurations preserving $1/4$  \cite{ggt} and $3/4$ \cite{gah} of the centrally extended  supersymmetries. The possibility for preserving  $3/4$  supersymmetries was earlier noted \cite{balu} to be one of the solutions for the superparticle \cite{ruds} moving in a  superspace extended by the coordinates of the tensorial central charge.
In  \cite{gght}  the combinations of momentum and domain-wall charges corresponding to  BPS  states preserving $1/2,  1/4$  and $3/4$ of  $D=4$ $ N=1$ supersymmetry  were constructed and it was  proved was that the Wess-Zumino model  does not admit any classical configurations with $3/4$ supersymmetry. 

A unified geometric approach to a description of superbranes was developed in 
\cite{caib} where new reasons for an enlargement of superspace were given. 

Here we propose new models for strings moving in $D=4$ space-time extended by
six real coordinates  $z_{mn}$  corresponding to the tensor central charges $Z_{mn}$.
The action  in our model is a natural generalization of a new  twistor like representation  \cite{gz}  for the Nambu-Goto and tensionless string actions. The suggested model admits  $N=1$ $D=4$ supersymmetrization and therefore may be effectively treated as a bosonic sector of superstrings moving in the generalized $D=(4+6)-$dimensional space-time. 

To make the dynamical role of the $z_{mn}$ coordinates  clear the simplest case of the tensionless string in a twistor-like formulation \cite{gz} is studied here.
 We find that even a minimal inclusion of the  $z_{mn}$ coordinates, linear in their derivatives, lifts  the light-like character of the tensionless string worldsheet and removes the degeneracy of the worldsheet metric. This is a hint  of string tension being generated. We solve the string equations together with the  system of integrability conditions and find a solution for  $x^{m}$ and  $z^{mn}$  which can be interpreated as a solution of the wave equation  $\mbox{\"X}_{\cal M}-X_{\cal M}^{\prime\prime}=0 $ describing a string in the extended D=10 space with the coordinates  $X^{\cal M}=(x^{m},z^{mn})$. In this solution the  $x^{m}$ coordinates have no transverse oscillations.
The solution for the  $z^{mn}$ coordinates implies the appearance of spin structure distributed along the string and a spin wave  \cite{vz},\cite{vz2} process related to this structure.
A supersymmetric version of the proposed string model is formulated.

\section {String action in $D=4$ space-time with\\ 
tensor central charge coordinates}

To describe the string dynamics we start from a twistor-like representation of the tensile/tensionless string action \cite{gz}

\begin{equation}\label{1}
S=\k\int (p_{mn}\,d{x^m}{\wedge}\,d{x^n}+ \L),
\end{equation}
which includes a local bivector $ p_{mn}(\t,\s)$ composed of the local Newman-Penrose dyads attached to the worldsheet and where the $\L-$term  fixes the orthonormality constraint  for the spinorial dyads (or twistor like variables). 

For the case of tensionless strings the Lagrange multiplier ${\L}$=0 and
$ p_{mn}(\t,\s)$ is  a null bivector defined by the condition 
\begin{equation}\label{2}
p_{mn}p^{mn}=0\quad,  \qquad
{\eta_{mn}=(-+++)},
\end{equation}
which implies 
\begin{eqnarray}\label{3}
p_{mn}(\t,\s)=i\bar{U}\g_{mn}U
\nonumber\\=2i[u^{\a}(\s_{mn})_{\a}^{\b}u^{\b} 
 + \bar{u}_{\dot\a}(\tilde{\s}_{mn})^{\dot\a}_{\dot\b}\bar{u}^{\dot\b}],
\end{eqnarray}
where $ U_{a}$  is a Majorana bispinor 
\begin{eqnarray}\label{4}
\nonumber
U_{a}=\left(\begin{array}[c]{c} u_{\a}\\ \bar{u}^{\dot\a} \end{array}\right) ,
\,\, \g_{mn}={1 \over 2}[\g_{m}, \g_{n}],\\ 
\s_{mn}={1 \over 4}(\s_{m}\tilde{\s_{n}} - \s_{n}\tilde{\s_{m}}).
\end{eqnarray}

For the  tensile string the bivector $ p_{mn}(\t,\s)$ may be represented as a sum of two null bivectors $p^{(+)}_{mn}$ and  $p^{(-)}_{mn}$    \cite{gz}
\begin{equation}\label{5}
p_{mn}=p^{(+)}_{mn} + p^{(-)}_{mn} =i\,[\bar{U}\g_{mn}U + \bar{V}\g_{mn}V],
\end{equation}
where $V_{a}=\left(\begin{array}[c]{c} v_{\a}\\ \bar{v}^{\dot\a} \end{array}\right)$ is  the second  component of the  Newman-Penrose dyads $(u_{\a}(\t,\s), v_{\a}(\t,\s))$
\begin{equation}\label{6}
u^{\a}v_{\a}=1 ,\quad u^{\a}u_{\a}=v^{\a}v_{\a}=0
\end{equation}
and the  $\L(\t,\s)-$term is 
\begin{equation}\label{7}
 \L(\t,\s)=\l(u^{\a}v_{\a}-1) -  \bar{\l}(\bar{u}^{\dot\a}\bar{v}_{\dot\a}-1).
\end{equation}

The action \f{1} may be rewritten in an equivalent spinor form
\begin{equation}\label{8}
S=i\k\int [\, p^{ab}\,d{x_{ae}}{\wedge}\,d{x_{db}}C^{ed}+ \L\, ],
\end{equation}
where $C^{ed}=(\g^0)^{ed}$  is the charge conjugation matrix in the Majorana
representation and  $p^{ab}$ is a symmetric local spin-tensor. In the general case $ p^{ab}$ may be represented as a bilinear combination of the Majorana bispinors $ U_{a}$  and  $V_{a}$.
\begin{equation}\label{9}
p^{ab}=\a\, U^{a}U^{b} + \b\, V^{a}V^{b} + \r\, (U^{a}V^{b} + U^{b}V^{a})
\end{equation}
with arbitrary coefficients  $\a,\b$ and $\r$.
 
The representation \f{8} includes an interesting object - the differential 2-form of the worldsheet area element $\xi_{a b}$ in the spinor representation
\begin{equation}\label{10}
\xi_{ab}=\xi_{ba}= C^{ed}\,d{x_{ae}}{\wedge}\,d{x_{db}},
\end{equation}
where
\begin{equation}\label{11}
dx_{ab}= (\g_{m}\,C^{-1})_{ab}\,d{x^{m}}
\end{equation}
 and $\xi_{ab}$ is a symmetric spin-tensor 2-form.

To include the real antisymmetric central charge coordinates $z_{mn}$
we consider the following  extension of $x_{ab}$
\begin{eqnarray}\label{13}
x_{ab}\longrightarrow Y_{ab}={x^{m}}\,(\g_{m}\,C^{-1})_{ab} 
\nonumber \\+iz_{mn}\,(\g^{mn}\,C^{-1})_{ab},
\end{eqnarray}
 used earlier in \cite{ruds} for the case of superparticles. As a result of \f{13} the area element $\xi_{ab}$ and the action \f{8} are replaced by 
\begin{equation}\label{14}
\xi_{ab}\longrightarrow \Xi_{ab}=dY_{al}\,{\wedge}{dY^{l}\,_{b}}.
\end{equation}
and 
\begin{eqnarray}\label{15}
S=i\k\int (p^{ab}d\,\Xi_{ab}+ \L\,) \nonumber \\
=i\k\int (p^{ab}\,dY_{ae}\,{\wedge}dY_{db}\,C^{ed}+ \L\,).
\end{eqnarray}

Our goal is to study the generalized action \f{15} and to this end we note that the generalized area element $\Xi_{ab}$  \f{14} is
\begin{eqnarray}\label{16}
\Xi_{a}\,^{b}=(\,d{x_m}{\wedge}\,d{x_n}-8dz_{ml}{\wedge}\,dz_{n}\,^{l})(\g^{mn})_{a}\,^{b}  \nonumber   \\
-4i\,d{x^l}{\wedge}\,dz_{lm}(\g^{m})_{a}\,^{b}.
\end{eqnarray}
Using properties of the $\g-$matrix algebra,
the action $S$ \f{15} takes the form
\begin{eqnarray}\label{19}
S=i\k\int \{\,[\,(\,dx_m{\wedge}\,dx_n - 8dz_{ml}{\wedge}\,dz_{n}\,^{l})\g^{mn}
 \nonumber  \\
-4i\,dx^l{\wedge}\,dz_{lm}\g^{m}]_{a}\,^{b}\,p^{a}\,_{b} +  \L\,\}.
\end{eqnarray}

In \f{19} $x_{m}$ and $z_{mn}$ appear on equal footing. As a first investigation of the model, however, we shall drop the quadratic $z-$term and study the  minimally extended action
\begin{eqnarray}\label{20}
S=i\k\int[(\,dx_m{\wedge}\,dx_n\g^{mn} \qquad\qquad \nonumber \\ \qquad 
-4i\,dx^l{\wedge}\,dz_{lm}\g^{m})_{a}\,^{b}\,p^{a}\,_{b}+\L],
 \end{eqnarray}
We may think of this action as the action  \f{19} in a certain limit  or as a model in its own right. In particular  we shall study the minimal extension of the tensionless string.

\section { Tensionless string in the space with \\
central charge coordinates $z_{\a\b}$} 

The action for the tensionless string \f{1}, \f{3} minimally extended by central charge coordinates is \f{20} with  $\L=0$ and $p^{ab}$ on the form \cite{gz}
\begin{equation}\label{21}
\nonumber
p^{ab}=U^{a}U^{b}.
\end{equation}
It reads
\begin{eqnarray}\label{22}
S=i\int U^{a}(\,dx_m{\wedge}dx_{n}\g^{mn} \qquad\qquad \nonumber \\ \qquad \qquad 
-4idx^l{\wedge}dz_{lm}\g^{m})_{a}\,^{b}U_{b}, 
\end{eqnarray}
where  $\k$  is  included in a redefinition of $x_{m}$ and $z_{lm}$ to make all variables \f{22} dimensionless.

The dynamics generated by the action $S$ \f{22} will in general break the light-like character of the string worldsheet \cite{sch}, \cite{kl}.
To study the dynamics  $S$  \f{22} we use the Weyl representation for $z_{mn}$ , where
$S$  \f{22}  takes the form 
\begin{eqnarray}\label{25}
S=2i\int\{\,dx_m{\wedge}\,dx_n(\,u\s^{mn}u +\bar u{\tilde\s}^{mn}\bar{u}) \qquad \nonumber \\ \qquad   
-dx^l{\wedge}\,[u_{\a}dz^{\a\b}(\s_{l}\bar{u})_{\b}- \bar u_{\dot\a}d\bar z^{\dot\a \dot\b}(u\s^{l})_{\dot\b}\,]\}.
\end{eqnarray}
To analyse the $u-$equations of motion we define the 2-forms
\begin{eqnarray}\label{26}
{\S_\a}^\b \equiv d{x_m}{\wedge}\,d{x_n}(\s^{mn})_{\a}\,^{\b}={1\over 2}\,
dx_{\a\dot\l}\wedge\,d\tilde x^{\dot\l\b}\nonumber \\
\tilde{\S}^{\dot\b}\,_{\dot\a}\equiv -({\S_\a}^\b)^\ast=
d{x_m}\wedge\,dx_{n}(\tilde{\s}^{mn})^{\dot\b}\,_{\dot\a}\qquad \nonumber\\
 ={1\over 2}\,d\tilde{x}^{\dot\b\l}\wedge dx_{\l\dot\a},\qquad \qquad 
\end{eqnarray}
and the antihermitian 2-form $\Om_{\a\dot\b}$ 
\begin{eqnarray}\label{28}
\Om_{\a\dot\b} \equiv -8i(dz_{ml}\wedge\,dx^{l})(\s^m)_{\a\dot\b}\nonumber \\
=2[(dz_{\a}\,^{\l}\wedge\,dx_{\l\dot\b}+dx_{\a\dot\l} d\bar z^{\dot\l}\,_{\dot\b}] , \nonumber \\
(\Om_{\a\dot\b})^\ast=-\Om_{\b\dot\a}.\qquad\qquad\qquad
\end{eqnarray}
Then the action  \f{25} becomes
\begin{equation}\label{29}
S=2i\int(u^{\a}{\S_\a}^{\b}u_{\b}+\bar{u}_{\dot\b}\tilde{\S}^{\dot\b}\,_{\dot\a}\bar{u}^{\dot\a}+{1\over 2}u^{\a}\Om_{\a\dot\b}\bar{u}^{\dot\b})
\end{equation}
and the equations of  motion for the dyad $ u_{\a}$ are
\begin{eqnarray}\label{30}
\nonumber 
{\S_\a}^{\b}u_{\b}+ {1\over 4}\Om_{\a\dot\b} \bar{u}^{\dot\b}=0, \\
\bar{u}_{\dot\b}\tilde{\S}^{\dot\b}\,_{\dot\a} +{1\over 4}u^{\b}\Om_{\b\dot\a}=0.
\end{eqnarray}
Under the assumption that  $ {\S_\a}^{\b}$ is non-degenerate \f{30} is equivalent to 
\begin{equation}\label{31}
({\S_\a}^{\b}- {1\over 16}{V_\a}^{\b})u_{\b}=0,
\end{equation}
where the  2-form ${V_\a}^{\l}\e_{\l\b}$ is the symmetric traceless 2-form
\begin{eqnarray}\label{32}
{V_\a}^{\b} \equiv (\Om\tilde{\S}^{-1}\e\Om^{\ast}\e)_{\a}\,^{\b}, \nonumber \\
{V_\a}^{\l}\e_{\l\b}={V_\b}^{\l}\e_{\l\a}.\qquad
\end{eqnarray}
Eqs. \f{30} are equivalent to the equation
\begin{equation}\label{33}
{\S_\a}^{\b}- {1\over 16}V_{\a}^{\b}=-Qu_{\a}u^{\b},
\end{equation}
where $Q$ is an arbitrary 2-form. All the matrices in Eq. \f{33} are  symmetric and their  determinants are  given by the relation
\begin{equation}\label{34}
\det{(\cal A_{\a}\,^{\b})}=-{1\over 2}Tr({\cal A}^2).
\end{equation}
Using \f{34} Eqs. \f{33} yield
\begin{equation}\label{35}
\det{\S}= ({1\over16})^{2}\det{V}+{1\over16}Qu_{\a}V^{\a}_{\b}u^{\b}.
\end{equation}
The definition \f{26} of ${\S_\a}^{\b}$ gives 
\begin{equation}\label{36}
\det{\S}={1\over2}(\,d{x^m}{\wedge}\,d{x^n})(\,d{x_m}{\wedge}\,d{x_n})
\end{equation}
and we conclude that the induced worldsheet metric is not in general  degenerate, i.e.
\begin{equation}\label{37}
\det{\S}\neq 0.
\end{equation}

\section { Solution  of the equations of motion}

To analyse the total set of string  equations we start from the Weyl representation for $S$ \f{25}
\begin{eqnarray}\label{38}
S=i\int[ (u^{\a}dx_{\a\dot\l}{\wedge}d\tilde x^{\dot\l\b}u_{\b}+ \bar{u}_{\dot\a}d\tilde {x}^{\dot{\a}\l}{\wedge}dx_{\l\dot\b}\bar{u}^{\dot\b})
\nonumber\\ 
+2 (u_{\a}dz^{\a\b}{\wedge}dx_{\b\dot\l}\bar{u}^{\dot\l} - \bar{u}_{\dot\a}d\tilde z^{\dot\a\dot\b}\wedge u^{\l}dx_{\l\dot\b}) ], 
\end{eqnarray}

The variation of  $S$ in \f{38} with respect to $z^{\dot\a\dot\b}$ and $x_{\a\dot\b}$ gives 
\begin{eqnarray}\label{39}
d\wedge\left(u_{\a}(dx\bar u)_{\b} +u_{\b}(dx\bar u)_{\a} \right)=0
\end{eqnarray}
and 
\begin{eqnarray}\label{40}
d\wedge[ -u^{\a}(d\tilde{x}u)^{\dot\b}+\bar u^{\dot\b}(\bar{u}d\tilde x)^{\a}\nonumber \\
+\bar u^{\dot\b}(udz)^{\a}-u^{\a}(ud\bar z)^{\dot\b} ]=0, 
\end{eqnarray}
respectively. (For simplicity we consider closed strings.) 
To solve Eqs. \f{39}, \f{40} and  \f{30}  we will use Cartan's method  (applied in \cite{zi,gz} for the solution of string dynamics). To this end recall that the spinors $u_{\a}$, $v_{\a}$ and $\bar u_{\dot\a}$,  $\bar v_{\dot\a}$ form a local spinor frame moving along string's worldsheet which may be used to build a local vector frame  with  
\begin{equation}\label{41}
u_{\a}\bar u_{\dot\a}, \;\, u_{\a}\bar v_{\dot\a}+v_{\a}\bar u_{\dot\a},\;\, 
v_{\a}\bar v_{\dot\a}, \;\, i( u_{\a}\bar v_{\dot\a}- v_{\a}\bar u_{\dot\a} )
\end{equation}
as basis elements. The independent differentials $ dx_{\a\dot\a}$  and  $dz_{\a\b}$ may be expanded in these basis elements as follows:
\begin{eqnarray}\label{42}
dx_{\a\dot\a}&=&x^{(u)}u_{\a}\bar u_{\dot\a}+
x^{(v)}v_{\a}\bar v_{\dot\a} \nonumber\\
&&+ x^{(+)}(u_{\a}\bar v_{\dot\a}+v_{\a}\bar u_{\dot\a})\nonumber \\
&&+ix^{(-)}(u_{\a}\bar v_{\dot\a} - v_{\a}\bar u_{\dot\a}), \nonumber \\
dz_{\a\b}&=&\vs^{(u)}u_{\a}u_{\b}+\vs^{(v)}v_{\a}v_{\b}\nonumber \\
&&+\vs du_{\a}v_{\b}+u_{\b}v_{\a}),\quad\quad \nonumber\\                    
d\bar z_{\dot\a\dot\b}&=&\bar\vs^{(u)}\bar u_{\dot\a}\bar u_{\dot\b}
+\bar\vs^{(v)}\bar v_{\dot\a}\bar v_{\dot\b} \nonumber\\
&&+\bar{\vs}( u_{\dot\a}\bar v_{\dot\b}+ u_{\dot\b}\bar v_{\dot\a})
\end{eqnarray}
(this defines $x^{(u)}, x^{v)}, x^{(+)}, x^{(-)},\vs^{(u)},\vs^{(v)}$ and  $\vs$).
Substituting the expansions \f{42} into Eqs. \f{30} we get a system of equations
\begin{eqnarray}\label{43}
x^{(v)}\wedge\,x^{(+)}&=&0,\nonumber \\
x^{(v)}\wedge(x^{(-)}-\vs_{I})
\nonumber\\ - \vs^{(v)}_{I}\wedge(x^{(+)}+x^{(-)})&=&0,
\nonumber\\
 x^{(u)}\wedge x^{(v)}
- 2\vs_R\wedge x^{(+)}\nonumber\\ 
+\vs^{(u)}_{R}\wedge\,x^{(v)}
+ \vs^{(v)}_{R}\wedge\,x^{(u)}&=&0,
\nonumber \\
2x^{(-)}\wedge\,x^{(+)}- \vs^{(u)}_{I}\wedge\,x^{(v)}
\nonumber \\ 
+\vs^{(v)}_{I}\wedge\,x^{(u)}
+2\vs_{R}{\wedge}\,x^{(-)}&=&0,
\end{eqnarray}
where $R$ and $I$ denote the  real and imaginary part, respectively. 
The first equation in \f{43} has the general solution 
\begin{equation}\label{45}
x^{(+)}=\l^{(+)}x^{(v)}.
\end{equation}
To solve the remaining equations of the system \f{43} we make a partial gauge fixing 
\begin{equation}\label{46}
\l^{(+)}=0  \; \Rightarrow  x^{(+)}(d)=0 , \; \; x^{(-)}(d)=0.
\end{equation}
The gauge \f{46} means  that the light-like vectors $u_{\a}\bar u_{\dot\a}$ and  $v_{\a}\bar v_{\dot\a}$ of the vector tetrad  are tangent vectors  to the string's worlsheet. As a result of this gauge choice, the $SO(3,1)$ local symmetry
 group of the vector frame \f{41} is reduced to its  $SO(1,1)$x$SO(2)$ subgroup and the expansion \f{42} for $dx_{\a \dot\a}$ simplifies to 
\begin{equation}\label{47}
dx_{\a\dot\a}=x^{(u)}(d)u{_\a}\bar u_{\dot\a}+x^{(v)}(d)v{_\a}\bar v_{\dot\a}.
\end{equation}
and Eqs.\f{43} become 
\begin{eqnarray}\label{48}
\nonumber  
x^{(v)}\wedge\,\vs_{I}=0, \\
\nonumber  
x^{(u)}\wedge\,x^{(v)}+\vs^{(u)}_{R}\wedge\,x^{(v)}+\vs^{(v)}_{R}\wedge\,x^{(u)} =0, \\
x^{(u)}\wedge\,\vs^{(v)}_{I}- x^{(v)}\wedge\,\vs^{(u)}_{I}=0.
\end{eqnarray}
In the gauge \f{46} the 2-forms  ${\S_\a}^{\b}$ and $\tilde{\S}^{\dot\b}\,_{\dot\a}$ \f{26} and $\det({\S_\a}^{\b})$ take the form
\begin{eqnarray}\label{50}
{\S_\a}^{\b}&=&-{1\over2}(u_{\a}v^{\b}+v_{\a}u^{\b})x^{(u)}\wedge x^{(v)},
\nonumber  \\
\tilde{\S}^{\dot\b}_{\dot\a}&=&{1\over2}(\bar{u}^{\dot\b}\bar{v}_{\dot\a} +
\bar{v}^{\dot\b}\bar{u}_{\dot\a})x^{(u)}\wedge x^{(v)}=0,
\nonumber \\
\det({\S_\a}^{\b})&=&\det( \partial_\mu x^{m}\partial_\nu x_{m})(d\t\wedge d\s)^2\nonumber \\ &=&-{1\over4}(x^{(u)}\wedge\,x^{(v)})^2 \nonumber\\
&=& -{1\over4}(\vs^{(u)}_{R}\wedge x^{(v)}+\vs^{(v)}_{R}\wedge x^{(u)})^2 
\end{eqnarray}

As  a result of further analysis of the equations  \f{48} we are able to 
 rewrite the expansions \f{42} as
\begin{eqnarray}\label{53}
dx_{\a\dot\a}&&=x^{(u)}(d)u_{\a}\bar u_{\dot\a}+
x^{(v)}(d)v_{\a}\bar v_{\dot\a},\nonumber\\
 dz_{\a\b}&&=(ax^{(u)}+bx^{(v)})u_{\a}u_{\b}\nonumber\\
+&&\left[cx^{(u)}+(1+\bar a)x^{(v)}\right]v_{\a}v_{\b}\nonumber\\
+&&(f_{R}x^{(u)}+g_{R}x^{(v)})(u_{\a}v_{\b}+u_{\b}v_{\a}),
\end{eqnarray}
where
\begin{eqnarray}\label{51}
\nonumber
\vs^{(u)}&=&ax^{(u)}+ bx^{(v)},\\
 a&=&a_{R}+ia_{I},\quad b=b_{R}+ib_{I},\nonumber \\
\vs^{(v)}&=&cx^{(u)}+ dx^{(v)},\nonumber\\
 c&=&c_{R}+ic_{I},\quad d=1+ \bar a,\nonumber \\
\vs_{R}&=&f_{R}x^{(u)}+ g_{R}x^{(v)},\nonumber \\
 f_{I}&=&0,\quad \vs_{I}=g_{I}x^{(v)}.
\end{eqnarray}

To solve the equations $\d S / \d {z_{\a\b}}=0$  \f{39} we expand  $du_{\a}$  and $dv_{\a}$ in the dyad basis
\begin{eqnarray}\label{52}
du_{\a}=\vp^{(u)}(d)u_{\a}+\vp^{(v)}(d)v_{\a},\nonumber\\
dv_{\a}=\psi^{(u)}(d)u_{\a}+\psi^{(v)}(d)v_{\a} 
\end{eqnarray}
and substitute  \f{52} into Eqs.\f{42}
\begin{equation}\label{54}
 \left [du_{\a}{\wedge}(dx\bar u)_{\b}-u_{\a}(dx{\wedge}d\bar u)_{\b}\right]+(\a \leftrightarrow \b) = 0.
\end{equation}
As a result we find the equations
\begin{eqnarray}\label{56}
(\vp^{(u)}+\bar\vp^{(u)})\wedge x^{(v)}=0 ,\nonumber\\
\vp^{(v)}\wedge x^{(u)}=\vp^{(v)})\wedge x^{(v)}=0,
\end{eqnarray}
which have the following general solution
\begin{eqnarray}\label{57}
Re\,\vp^{(u)}&\equiv&\vp^{(u)}_{R}=\a^{(u)}_{R}x^{(v)} ,\nonumber\\
\vp^{(v)}(d)&=&0
\end{eqnarray}
with $\a^{(u)}_{R}(\t,\s)$  an arbitrary real function.

We now turn to the equations $\d S / \d {x_{\a\dot\b}}=0$ in \f{43}.
Using  the solutions of Eqs. $\d S / \d u_{\a}=0$ in  \f{53} and of
$\d S / \d z_{\a\b}=0$ in \f{57} it may be written
\begin{eqnarray}\label{59}
\nonumber
(\vp^{(u)}_{I}+\a^{(u)}_{R}c_{I}x^{(u)}){\wedge}x^{(v)}&=&0,\\
\a^{(u)}_{R}c_{R}x^{(v)}{\wedge}x^{(u)}&=&0
\end{eqnarray}
Eqs.  \f{59} have two sets of solutions. The first  set is 
\begin{eqnarray}\label{60}
\vp^{(u)}_{I}&=&\a^{(u)}_{I}x^{(v)},\nonumber\\
\a^{(u)}_{R}&=&0
\end{eqnarray}
 and the second set is 
\begin{eqnarray}\label{61}
\vp^{(u)}_{I}&=&-\a^{(u)}_{R}c_{I}x^{(u)} +\a^{(u)}_{I}x^{(v)},\nonumber\\
c_{R}&=&0 ,
\end{eqnarray}
where $\a^{(u)}_{I}(\t,\s)$ is an arbitrary function.

Having found  the general solutions \f{53}, \f{57} and \f{60} or \f{61} to the equations of motion we  now also have  to analyse the integrability conditions for the expansions \f{53} and \f{54}. These integrability conditions will play the role of dynamical equations for the string.

\section {Solution of the integrability conditions\\ for the  $dx_{\a\dot\a}$, 
$du_{\a}$ and $dv_{\a}$ expansions}


The integrability conditions $(IC)$  $d{\wedge}dx_{\a\dot\a}=0$  for the $ dx_{\a\dot\a}-$ expansion  \f{53} are 
\begin{eqnarray}\label{63}
 d{\wedge}x^{(u)}+2\a^{(u)}_{R}x^{(v)}{\wedge}x^{(u)} =0, \nonumber\\
d{\wedge}x^{(v)}+ 2\psi^{(v)}_{R}{\wedge}x^{(v)}=0, \nonumber \\
 x^{(v)}{\wedge}\bar\psi^{(u)}=0,\nonumber \\
x^{(v)}{\wedge}\psi^{(u)}=0,
\end{eqnarray}
where we have used \f{60}.
It follows directly that 
\begin{equation}\label{64}
\psi^{(u)}=\tilde\Th^{(u)} x^{(v)},
\end{equation}
where $\tilde\Th^{u}$ is an arbitrary function
 The  $IC$ for the  $(du,dv)-$ differential expansions \f{54} may be written as  \begin{eqnarray}\label{66}
 d{\wedge}\vp^{(u)}_{I}=0, \nonumber\\
d{\wedge}\vp^{(u)}_{R}\equiv d{\wedge}( \a^{(u)}_{R}x^{(v)})=0, \nonumber\\
 d{\wedge}\psi^{(v)}=0, \nonumber\\
d{\wedge}(\tilde\Th^{(u)}x^{(v)}+ (i\vp^{(u)}_{I}-\psi^{(v)}){\wedge}\tilde\Th^{(u)}x^{(v)}=0,
\end{eqnarray}
where  $ \tilde\Th^{(u)}$ is 
\begin{equation}\label{71}
\tilde\Th^{(u)}=e^{-(i\tilde\a^{(u)}_{I} -\tilde\Th^{(v)})}\Th^{(u)}(\t,\s)
\end{equation}

The two sets of equations $IC$ \f{63} and  \f{66} may be combined into the relations
\begin{eqnarray}\label{72}
\vp^{(u)}_{I}=d\tilde\a^{(u)}_{I},\;  \vp^{(v)}=0,\nonumber \\
\vp^{(u)}_{R}=\a^{(u)}_{R}x^{(v)}=d\tilde\a^{(u)}_{R}, \nonumber \\
\psi^{(u)}=e^{-(i\tilde\a^{(u)}_{I} -\tilde\Th^{(v)})} \Th^{(u)} x^{(v)},
\nonumber \\
 \psi^{(v)}=d\tilde\Th^{(v)}
\end{eqnarray}
and into the simple system 
\begin{eqnarray}\label{73}
d{\wedge}(e^{2\tilde\a^{(u)}_{R}}x^{(u)})=0,   \nonumber\\
d{\wedge}(\e^{2\tilde\Th^{(v)}_{R}}x^{(v)})=0,\nonumber\\
d{\wedge}(\Th^{(u)}x^{(v)})=0,
\end{eqnarray}
where  $\tilde\a^{(u)}_{I}(\t,\s)$ ,$\tilde\Th^{(v)}(\t,\s)$ are arbitrary functions. 

In addition the integrability conditions
\begin{equation}\label{74}
d{\wedge}dz_{\a\b}=0,
\end{equation}
needs to be analysed and the solutions \f{60} and \f{61} should be taken into account.

\section {Solution of the integrability conditions\\ for the  $dz_{\a\b}$ expansion}

If we use  \f{51}, \f{52}, \f{57} and  \f{72} the $dz-$  $IC$'s may be written
\begin{eqnarray}\label{78}
d{\wedge}\left(e^{2\tilde\a^{(u)}}ax^{(u)}+e^{2\tilde\a^{(u)}}bx^{(v)}\right)
+\nonumber \\
2e^{(2\tilde\a^{(u)}+ i\tilde\a^{(u)}_{I}+ \tilde\Th^{(v)})}\Th^{(u)}f_{R}x^{(v)}{\wedge}x^{(u)}&=&0, \nonumber \\ 
d{\wedge}\left(e^{2\tilde\Th^{(v)}}cx^{(u)}+e^{2\tilde\Th^{(v)}}(1+\bar a)x^{(v)}\right)&=&0, \nonumber \\
d{\wedge}\left(e^{2(\tilde\a^{(u)}+\tilde\Th^{(v)})}[f_{R}x^{(u)}+gx^{(v)}]\right) + \nonumber \\
e^{(2\tilde\a^{(u)}_{R}+ i\tilde\a^{(u)}_{I}+ 3\tilde\Th^{(v)})}\Th^{(u)}cx^{(v)}{\wedge}x^{(u)}&=&0 
\end{eqnarray}

Thus, the final part of our analysis concerns  the solution of the equatioqns \f{78} and \f{73} and  together with the solutions \f{63} or \f{64}. Here we shall solve Eqs. \f{78} and \f{73} choosing the solution \f{60} which prescribes that
\begin{eqnarray}\label{79}
\a^{(u)}_{R}&=&\tilde\a^{(u)}_{R}=0, \nonumber \\ 
 \vp^{(u)}_{I}&=&\a^{(u)}_{I}x^{(v)}= d\tilde\a^{(u)}_{I} .
\end{eqnarray}
In that case  Eqs.\f{73} reduces to the equations
\begin{eqnarray}\label{80}
d{\wedge}x^{(u)}&=&0, \nonumber \\     
d{\wedge}(e^{2\tilde\Th^{(v)}_{R}}x^{(v)})&=&0,  \nonumber \\     
d{\wedge}(\Th^{(u)}x^{(v)})&=&0.
\end{eqnarray}

Returning to $IC$ \f{78} we note that the last equation may be satisfied if 
the arbitrary functions $f_{R}$,  $g$, $\tilde\Th^{(v)}_{I}$, $c\Th^{(u)}$  are fixed by  the relations
\begin{eqnarray}\label{81}
e^{2\tilde\Th^{(v)}_{R}}f_{R}={\cal A}_{f}=const, \nonumber \\ 
\tilde\Th^{(v)}_{I}+ \tilde\a^{(u)}_{I}=\triangle_{0}=const, \nonumber \\ 
g_{R}={\cal A}_{R}=const, \;  g_{I}={\cal A}_{I}=const,   \nonumber \\ 
c\Th^{(u)}=0.
\end{eqnarray}
The last  equation in \f{81} has two solutions
\begin{equation}\label{82}
\Th^{(u)}=0 \Longrightarrow \psi^{(u)}=0
\end{equation}
and 
\begin{equation}\label{83}
c=0.
\end{equation}

The case \f{82} seems to be the simplest for further analysis.
In that case  the $IC$ \f{78} are reduced to the form
\begin{eqnarray}\label{84}
\Th^{(u)}=\psi^{(u)}=0,   \nonumber \\   
d{\wedge}\left( e^{2i\tilde\a^{(u)}_{I}}[ax^{(u)}+ bx^{(v)}]\right)=0 ,\nonumber \\     
d{\wedge}\left( e^{2(\tilde\Th^{(v)}_{R}-i\tilde\a^{(u)}_{I})}[cx^{(u)}
+(1+\bar a)x^{(v)}] \right)=0.
\end{eqnarray}
Still there is a possibiulity to  furter simplify  the reduced  $IC$  \f{84} fixing an arbitrariness in the definitions of the functions $a$,  $b$  and  $c$ by the relations
\begin{eqnarray}\label{85}
b=b_{0}\,e^{2(\tilde\Th^{(v)}_{R}-i\tilde\a^{(u)}_{I})}, \;\; b_{0}=const. , 
\nonumber \\  
 a=a_{0}e^{-2i\tilde\a^{(u)}_{I}},  \; \; a_{0}=\bar a_{0}=const, \nonumber \\
c=c_{0}\,e^{-2(\tilde\Th^{(v)}_{R}-i\tilde\a^{(u)}_{I})},\;\; c_{0}=const.  
\end{eqnarray}
 Then, as a result of  \f{80}, we find the single  $IC$
\begin{equation}\label{86}
d\tilde\a^{(u)}_{I}{\wedge}x^{(v)}=0,
\end{equation}
which is identically satisfied using the second relation in equation \f{79}.

Thus, the total set of the $IC$'s under consideration is reduced to the set
\begin{eqnarray}\label{87}
d{\wedge}x^{(u)}&=&0, \nonumber \\     
d{\wedge}(e^{2\tilde\Th^{(v)}_{R}}x^{(v)})&=&0,  \nonumber \\     
d\tilde\a^{(u)}_{I}&=&\a^{(u)}_{I}x^{(v)},
\end{eqnarray}
which has the following solutions
\begin{eqnarray}\label{88}
x^{(u)}=d\eta^{(u)} , \nonumber \\     
x^{(v)}=e^{-2\tilde\Th^{(v)}_{R}}d\eta^{(v)}, \nonumber \\   
d\tilde\a^{(u)}_{I}=\a^{(u)}_{I}e^{-2\tilde\Th^{(v)}_{R}}d\eta^{(v)},
\end{eqnarray}
where $\eta^{(u)}(\t,\s)$  and $\eta^{(v)}(\t,\s)$ are atbitrary functions.
The last $IC$  in \f{88} is easily satisfied  choosing 
\begin{equation}\label{89}
\a^{(u)}_{I}=e^{2\tilde\Th^{(v)}_{R}}\b^{\prime}(\eta^{(v)})
\end{equation}
which gives the following representation for $\tilde\a^{(u)}_{I}$
\begin{equation}\label{90}
\tilde\a^{(u)}_{I}=\b(\eta^{(v)}(\t,\s)),
\end{equation}
where $\b(\eta^{(v)})$ is an arbitrary functions. 

In the next section we shall  discuss an example of string motion arising from the above  equations.

\section {An example of the string motion\\in the extended space-time}

To get an example of string motion let us fix the function $\b(\eta^{(v)})$ in the solution \f{90} by the condition  $\b^{\prime}(\eta^{(v)})=1$ which gives 
\begin{equation}\label{91}
\tilde\a^{(u)}_{I}=\eta^{(v)},\;\; \a^{(u)}_{I}=\a_{0}e^{2\tilde\Th^{(v)}_{R}}
\end{equation}
and then the total set of the integrability conditions is satisfied.

 The final equations \f{53} corresponding to the solutions \f{91} and \f{88} and describing the string dynamics take the simple form
\begin{eqnarray}\label{92}
dx_{\a\dot\a}=u_{\a}\bar u_{\dot\a}d\eta^{(u)}+
v_{\a}\bar v_{\dot\a}e^{-2\tilde\Th^{(v)}_{R}}d\eta^{(v)}, \nonumber\\
 dz_{\a\b}=e^{-2i\eta^{(v)}}(a_{0}d\eta^{(u)}+b_{0}d\eta^{(v)})u_{\a}u_{\b} \nonumber\\+
e^{-2\tilde\Th^{(v)}_{R}}[ c_{0}e^{2i\eta^{(v)}}d\eta^{(u)}
\nonumber\\
+(1+a_{0}e^{2i\eta^{(v)}})d\eta^{(v)}]v_{\a}v_{\b}\nonumber\\
+e^{-2\tilde\Th^{(v)}_{R}}({\cal A}_{f}d\eta^{(u)}+{\cal A}d\eta^{(v)}
)(u_{\a}v_{\b}+u_{\b}v_{\a}), \nonumber\\
du_{\a}=id\eta^{(v)}u_{\a},\nonumber\\
dv_{\a}=(-id\eta^{(v)}+d\tilde\Th^{(v)}_{R})v_{\a}.
\end{eqnarray}

To analyse Eqs.\f{92} we note that the spinor subset has the general solution \begin{eqnarray}\label{93}
u_{\a}=u_{0\a}e^{i\eta^{(v)}(\t,\s)} , \nonumber\\
v_{\a}=v_{0\a}e^{-i\eta^{(v)}(\t,\s)+\tilde\Th^{(v)}_{R}(\t,\s)},
\end{eqnarray}
where $u_{0\a}$ and  $v_{0\a}$ are arbitrary constant spinors.\\
Substituting the solutions \f{93} into the constraints \f{6} results in the same constraints for both dyads   $u_{0}^{\a}$  and   $v_{0}^{\a}$
\begin{equation}\label{94}
u_{0}^{\a}v_{0\a}=1, \,\quad u_{0}^{\a}u_{0\a}=v_{0}^{\a}v_{0\a}=0,
\end{equation}
and  implies
\begin{equation}\label{95}
 \tilde\Th^{(v)}_{R}=0.
\end{equation}

So, we find that for  the presented example the whole dynamics of the dyads $u_{\a}$ and $v_{\a}$ is reduced to phase transformations
\begin{eqnarray}\label{96}
u_{\a}=u_{0\a}e^{i\eta^{(v)}(\t,\s)} , \nonumber\\
v_{\a}=v_{0\a}e^{-i\eta^{(v)}(\t,\s)}.
\end{eqnarray}
It follows from the solutions \f{96} that 
\begin{eqnarray}\label{97}
u_{\a}\bar u_{\dot\a}&=&u_{0\a}\bar u_{0\dot\a}=const,  \nonumber\\
v_{\a}\bar v_{\dot\a}&=&v_{0\a}\bar v_{0\dot\a}=const.
\end{eqnarray}

The conditions \f{97} essentially simplify equation \f{92} for the string world vector  $x_{\a\dot\a}$ transforming it into the equation
\begin{equation}\label{98}
dx_{\a\dot\a}=d\eta^{(u)}u_{0\a}\bar u_{0\dot\a}+
d\eta^{(v)}v_{0\a}\bar v_{0\dot\a}
\end{equation}
which has the following general solution 
\begin{equation}\label{99}
x_{\a\dot\a}=x_{0\a\dot\a}+
\eta^{(u)}u_{0\a}\bar u_{0\dot\a}+
\eta^{(v)}v_{0\a}\bar v_{0\dot\a},
\end{equation}
where $\eta^{(u)}$ and $\eta^{(v)}$ are arbitrary functions on the worldsheet.
According to Eqs.\f{88}, which  now take the form
\begin{equation}\label{100}
x^{(u)}=d\eta^{(u)},\,\;  x^{(v)}=d\eta^{(v)},
\end{equation}
one finds from  \f{50} and  \f{100}   that
\begin{eqnarray}\label{101}
\det({\S_\a}^{\b})=\det( \partial_\mu x^{m}\partial_\nu x_{m})(d\t\wedge d\s)^2\nonumber \\ 
=-{1\over4}(x^{(u)}\wedge\,x^{(v)})^2 = 
-{1\over4}(d\eta^{(u)}\wedge\,d\eta^{(v)})^2,
\end{eqnarray}
so that the expected condition \f{37} $\det{\S}\neq 0$, is satisfied for the solution \f{99} and the induced  worldsheet metric of the tensionless string becomes regular.
As the vectors 
\begin{eqnarray}\label{102}
{\partial x_{\a\dot\a} \over \partial \eta^{(u)}}=u_{0\a}\bar u_{0\dot\a}\equiv (\s_{m})_{\a \dot\a}A^{m},\nonumber\\
{\partial x_{\a\dot\a} \over \partial \eta^{(v)}}=v_{0\a}\bar v_{0\dot\a}\equiv (\s_{m})_{\a \dot\a}B^{m}
\end{eqnarray}
tangent to the worldsheet are constant light-like vectors we conclude that 
the worldsheet \f{10} lies in the plane spanned by the  light-like vectors $A^{m}$ and  $B^{m}$ \begin{eqnarray}\label{103}
 x^{m}=x_{0}^{m}+ A^{m}\eta^{(u)}(\t+\s)+ B^{m}\eta^{(v)}(\t-\s),\nonumber\\
A^{m}A_{m}=B^{m}B_{m}=0, A^{m}B_{m}=1. \qquad  
\end{eqnarray}
and we see that the central charge coordinates do not  excitate the transverse oscillations of the  $x^{m}$-coordinates. But what about the $z^{mn}$-coordinates?

By analogy with the previous case taking into account the solutions \f{93} simplifies Eqs.\f{92} for the central charge coordinates $z_{\a\b}$ to the form
\begin{eqnarray}\label{104}
dz_{\a\b}=(a_{0}d\eta^{(u)} + b_{0}d\eta^{(v)})u_{0\a}u_{0\b} \nonumber\\
+ \left( c_{0}d\eta^{(u)}+ (e^{-2i\eta^{(v)}}+ a_{0})d\eta^{(v)}\right )v_{0\a}v_{0\b}\nonumber\\
+ ({\cal A}_{f}d\eta^{(u)}+ {\cal A}d\eta^{(v)})(u_{0\a}v_{0\b}+u_{0\b}v_{0\a})
\end{eqnarray}
and the general solution of \f{104} is
\begin{eqnarray}\label{105}
z_{\a\b}=z_{0\a\b} + (a_{0}\eta^{(u)}+ b_{0}\eta^{(v)})u_{0\a}u_{0\b} \nonumber\\
+ ( c_{0}\eta^{(u)}+a_{0}\eta^{(v)} - 2ie^{-2i\eta^{(v)}} )v_{0\a}v_{0\b} \nonumber\\
+({\cal A}_{f}\eta^{(u)}+{\cal A}\eta^{(v)})(u_{0\a}v_{0\b}+u_{0\b}v_{0\a}).
\end{eqnarray}

The solutions \f{99} and \f{105} can be interpreted as solutions of the two-dimensional wave equations
\begin{eqnarray}\label{106} 
\mbox{\"x}_{\a\dot\a}-x^{\prime\prime}_{\a\dot\a}=0 ,   \nonumber \\
\mbox{\"z}_{\a\b}-z^{\prime\prime}_{\a\b}=0,
\end{eqnarray}
which may be presented in the form 
\begin{equation}\label{107}
\mbox{\"X}_{\cal M} - X_{\cal M}^{\prime\prime}=0, 
\end{equation}
 where $X^{\cal M}=(x^{m},z^{mn})$ are generalized coordinates in the extended
 D=10 space-time.
 The wave equation  \f{107} implies  that the coordinates $z^{mn}$ have 
an interpretation similar to that  for the $x^{m}$ namely as  additional string coordinates in the generalized  D=10 space.
After the substitution of   
the solution \f{105}  into the representation 
\begin{equation}\label{108}
z_{mn}={i\over 4}[z_{\a}\,^{\b}(\s_{mn})_{\b}\,^{\a}+\bar{z}_{\dot\a}\,^{\dot\b}(\tilde{\s}_{mn})^{\dot\a}\,_{\dot\b}]
\end{equation}
for $z^{mn}$ we find the  appearance of the spinorial structures 
$(u_0\s_{mn}u_0)$,
$(v_0\s_{mn}v_0)$ and $(u _0\s_{mn}v_0)$, which correspond to the spin degreees of freedom distributed along the string worldsheet. These spin factors are multiplied by the functions  which are  solutions of the wave equation \f{107}.
Therefore, we may understand the solution  \f{105} as describing a spin wave process associated with the $z^{mn}$ degrees of freedom.

\section {Conclusion}

 We have  suggested a new model for strings embedded  into $D=4$  space-time extended by  6 additional coordinates $z^{mn}$ corresponding to the tensor central charge $Z^{mn}$.
In studying the simplest case of a tensionless string we found that 
 presence of the  $z^{mn}$  coordinates lifts the degeneration of the worldsheet metric  typical for the tensionless string.
 We found a solution of the model which gives an example of string dynamics in the extended $D=(4+6$)-dimensional space. For this  solution  the  $x^{m}$-coordinates do not develop  transverse oscillations.
To understand if  this result is general or if the  $x^{m}$ coordinates also can oscillate we need to analyse the general solution to the integrability conditions derived here or/and to extend the minimal  model. On the other hand, the evolution of the $z^{mn}$  coordinates may be understood  as a spin wave process assosiated with an excitation of  spin degrees of freedom distributed along the string. 

The next step is to study the general case of the tensile string described by the general action $\f{15}$ with the spin-tensor  $p^{ab}$ given by \f{9}. 

The  supersymmetrized version of the model is found  by going from the differentials $dx^{m}$ and $dz^{mn}$ to the supersymmetric  invariant differential Cartan forms $\Pi^{m}$ and $\Pi^{mn}$ in the general representation \f{15} or \f{19}  and adding  a  Chern-Simons three-form \cite{curt}. Therefore, an  example of a superstring action in  $D=4$ space-time extended by the coordinates $z^{mn}$ is
\begin{eqnarray}\label{109}
\tilde S=i\k\int \{[(\Pi_m{\wedge}\Pi_n  \nonumber \\ - 8\Pi_{ml}{\wedge}\Pi_{n}\,^{l})\g^{mn}
-4ic\Pi^l{\wedge}\Pi_{lm}\g^{m}]_{a}\,^{b}p^{a}\,_{b} \nonumber \\
 + \L + (Chern-Simons) \},
\end{eqnarray}
where the constant $c$ may be zero. 
The extended superalgebra  with  tensorial central charges $Z^{mn}$ arising from the superstring action \f{108} (and/or its brane-generalizations)  may contain new information on the preserved supersymmetry. An investigation of this 
is in progress.

\section {Acknowledgements}

A.Z. thanks ITP at the Stockholm University for the kind hospitality.
The work is  partially supported by the grants of the Royal Swedish Academy of Sciences and Axel Wenner-Gren Foundation. The work of A.Z. is also partially  supported by the Award CRDF-RP1-2108. U.L. is supported in part by NFR grant 5102-20005711 and by EU contract HPRN-C7-2000-0122.


\begin{thebibliography}{99}

\bibitem{alb}
A. Schwarz, Gauge theories on nocommutative spaces, hep-th/0011261.
\bibitem{codos}
A. Connes, M. Douglas and A.S. Schwarz, \JHEP 02 1998 003; hep-th/9711162.
\bibitem{dohul}
 M. Douglas and C. Hull,
\JHEP 02 1998 008.
\bibitem{seiwi}
N. Seiberg and E. Witten,
 \JHEP 09 1999 032; hep-th/9908142.
\bibitem{seisut}
N. Seiberg, L.Susskind and N. Toumbas,\JHEP 0006 2000 021; hep-th/0005040.
\bibitem{gomms}
R. Gopakumar, J. Maldacena, S. Minwalla and A. Strominger,\JHEP 0006 2000 036; hep-th/0005048.
\bibitem{gomss}
R. Gopakumar, S. Minwalla  J. Maldacena and A. Strominger,\JHEP 0008 2000 008;  hep-th/0006062.
\bibitem{gukp}
S. Gukov, I.R. Klebanov,and A.M. Polyakov,
\PLB 423 1998 64.
\bibitem{sch}
A. Schild,  \PRD 16 1977 1722.
\bibitem{kl} 
A. Karlhede and U. Lindstr\"om, \CQG 3 1986 L73. 
\bibitem{zb}
A.A. Zheltukhin,\SJNP 48 1988 950.\\
I.A. Bandos and A.A. Zheltukhin,  \FdP 41 1993 619.
\bibitem{lir}
U. Lindstr\"om and R. von Unge,
 \PLB 403 1997 233; hep-th/9704051;\\
H. Gustafson and U. Lindstr\"om, , \PLB 440 1998 43; hep-th/9807064.
\bibitem{lizz}
U. Lindstr\"om, M. Zabzine and A.A. Zheltukhin,
 \JHEP 12 1999 016; hep-th/9910159.
\bibitem{zr}
A.A. Zheltukhin, \CQG 12 1996 2357, hep-th/9606013;\\
S.N.Roshchupkin and A.A. Zheltukhin,
 \NPB 543 1999 365, hep-th/9806054.
\bibitem{zi}
A.A. Zheltukhin, \SJNP 51 1990 950;\\
 K. Ilienko and A.A. Zheltukhin,
\CQG 16 1999 383.
\bibitem{haliu} 
S. Hassani, U. Lindstr\"om and R, von Unge, \CQG 11 1994 L79.
\bibitem{gz} 
O.E. Gusev and A.A. Zheltukhin,  \JETP 64 1996 487.
\bibitem{town1}
 P.K. Townsend,
\PLB 277 1992 285.
\bibitem{belt} 
E.Bergshoeff, L.A. London and P.K. Townsend, \CQG 9 1992 2545.
\bibitem{hupol}
J. Huges and J. Polchinski, \NPB 278 1986 147.
\bibitem{azgit}
J.A. de Azcarraga, J.P. Gauntlett, J.M. Izquierdo and  P.K. Townsend, 
\PRL 63 1989 2443.
\bibitem{dust}
M. Duff and K.S. Stelle,
 \PLB 253 1991 113.
\bibitem{dafre}
R. D'Auria and P. Fre, \NPB 201 1982 101.
\bibitem{holpro}
J. van Holten and A. van Proyen,
  \JPA 15 1982 3763.
\bibitem{ziz} 
P.A. Zizzi, \PLB 149 1984 333. 
\bibitem{curt}
T. Curtright,
\PRL 60 1988 393. \bibitem{gren}
M.B. Green,
 \PLB 223 1989 157.
\bibitem{hew}
 S. Hewson, \NPB 507 1997 445; hep-th/9701011.
\bibitem{eis}
Y. Eisenberg and S. Solomon,
 \PLB 220 1988 562.
\bibitem{sieg}
W. Siegel, \PRD 50 1994 2799.
\bibitem{bes}
 E. Bergshoeff and E.Sezgin,\PLB 354 1995 256.
\bibitem{se}
 E. Sezgin, \PLB 392 1997 323.
\bibitem{wit}
E. Witten,
 \NPB 443 1995 85.
\bibitem{town2}
 P.K. Townsend,  M-theory from its superalgebra; hep-th/9712004.
\bibitem{pol1}
 J. Polchinski, 
\PRL 75 1995 184.
\bibitem{ds} G. Dvali and M. Shifman, \PLB 396 1997  64; erratum \PLB 407 1997 452; hep-th/9612128.
\bibitem{gs}
A. Gorsky and  M. Shifman,  More on the tensor central charges in  N=1 supersymmetric gauge theories ( BPS wall junctions and strings),hep-th/9909015.  
\bibitem{vy}
G. Veneziano and S. Yankielowicz,\PLB 113 1982 231.
\bibitem{ggt}
 P.K. Townsend and G.W. Gibbons,\PRL 83 1999 1727.
\bibitem{gah}
 J.P. Gauntlett and C.M. Hull, \JHEP 001  2000 004; hep-th/9909098.
\bibitem{balu}
I. Bandos and J. Lukierski, \MPLA 14 1999 1257.
\bibitem{ruds}
I. Rudychev and E. Sezgin, \PLB 415 1997 363 ;  hep-th/9711128.
\bibitem{gght}
J.P. Gauntlett, G.W. Gibbons, C.M. Hull and  P.K. Townsend, \CMP 216 2001  431.
\bibitem{caib}
C. Chryssomalakos. J.A. de Azcarraga. J.M. Izquierdo, J.C. Perez Bueno, \NPB 567 2000 293. 
\bibitem{vz}
 D.V. Volkov and A.A. Zheltukhin, \SSC 36 1980 733.
\bibitem{vz2}
 D.V. Volkov and  A.A. Zheltukhin, \JETP 78 1980 1867.
\end{thebibliography}
\end{document}